\documentclass[aps,twocolumn,prr,superscriptaddress,floatfix]{revtex4-2}

\usepackage[utf8]{inputenc}
\usepackage[colorlinks=true]{hyperref}
\usepackage{amsmath}
\usepackage{amsfonts}
\usepackage{amssymb}
\usepackage{bm}
\usepackage{scalerel}
\usepackage{graphicx}
\usepackage{xcolor}

\newcommand{\avg}[1]{\langle{#1}\rangle}

\newcommand{\mc}{\mathcal}

\newcommand{\tbl}[1]{#1}

\begin{document}

\title{Nanomechanically-induced nonequilibrium quantum phase transition to a self-organized density wave of a Bose-Einstein condensate}

\author{Milan Radonji\'{c}}
\email{milan.radonjic@physik.uni-hamburg.de}
\affiliation{I. Institut f\"ur Theoretische Physik, Universit\"at Hamburg, Notkestr.~9, 22607 Hamburg, Germany}
\affiliation{Institute of Physics Belgrade, University of Belgrade, Pregrevica 118, 11080 Belgrade, Serbia}
\author{Leon Mixa}
\email{lmixa@physnet.uni-hamburg.de}
\affiliation{I. Institut f\"ur Theoretische Physik, Universit\"at Hamburg, Notkestr.~9, 22607 Hamburg, Germany}
\affiliation{The Hamburg Center for Ultrafast Imaging, Luruper Chaussee 149, 22761 Hamburg, Germany}

\author{Axel Pelster}
\email{axel.pelster@rptu.de}
\affiliation{Physics Department and Research Center OPTIMAS, University Kaiserslautern-Landau, Erwin-Schr\"odinger Str.~46, 67663 Kaiserslautern, Germany}
\author{Michael Thorwart}
\email{michael.thorwart@physik.uni-hamburg.de}
\affiliation{I. Institut f\"ur Theoretische Physik, Universit\"at Hamburg, Notkestr.~9, 22607 Hamburg, Germany}
\affiliation{The Hamburg Center for Ultrafast Imaging, Luruper Chaussee 149, 22761 Hamburg, Germany}

\begin{abstract}
We report on a nonequilibrium quantum phase transition (NQPT) in a hybrid quantum many-body system consisting of a vibrational mode of a damped nanomembrane interacting optomechanically with a cavity, whose output light couples to two internal states of an ultracold Bose gas held in an external quasi-one-dimensional box potential. For small effective membrane-atom couplings, the system is in a homogeneous Bose-Einstein
condensate (BEC) steady state, with no membrane displacement. Depending on the transition frequency between the two internal atomic states, either one or both internal states are occupied. By increasing the atom-membrane couplings, the system transitions to a symmetry-broken self-organized BEC phase, which is characterized by
a considerably displaced membrane steady-state and density-wave-like BEC profiles. This NQPT can be both discontinuous and continuous for a certain interval of transition frequencies and is purely discontinuous outside of it.
\end{abstract}

\date{\today}

\maketitle

\section{Introduction}

Many physical systems can be classified in terms of their emergent collective behavior when they undergo phase transitions \cite{Sachdev2011,Zinn-Justin2002,Kleinert2001}. While equilibrium phase transitions are fairly well understood, nonequilibrium quantum phase transitions (NQPTs) are a relatively new and exciting field of research \cite{review1,review2}. Such phase transitions occur in a nonequilibrium steady state of driven quantum systems in contact with an environment, so that their energy is not conserved due to external driving and dissipation \cite{Mitra2006,Mitra2007,Brandes2012,Brennecke2013,Marino2016,Gorshkov2016,Larson2018,Rota2019,Puel2019,Kennes2020,Behrle2023}. Among other cases, NQPTs have been studied in ultracold atoms in an optical cavity \cite{Nagy2008,Ritsch2008,Esslinger2010,Ritsch2013,Hemmerich2015,Thorwart2015,Ritsch2021,Defenu2023}, in driven-dissipative Bose-Einstein condensates \cite{Mink,Ott}, in microcavity-polariton systems \cite{Szymanska2015,Szymanska2017,Szymanska2018}, and in photon Bose-Einstein condensates \cite{Nyman2018,Nyman2021,Weitz}.
Additionally, dynamical quantum phase transitions and universal scaling have been observed both in the nonequilibrium dynamics of isolated quantum systems after a quench, with time playing the role of the control parameter \cite{Heyl}, and recently in open quantum systems, whose dynamics is driven by the dissipative contact to an environment \cite{Widera}.

Hybrid atom-optomechanical systems, which combine optomechanics with atom optics, represent one recent promising platform for studying NQPTs \cite{Treutlein2013,Vogell2015,Treutlein2015,Zhong2017,Becker2018,Treutlein2018,Mann2018,Gao2019,Treutlein2022b}. In these systems, a nanomembrane in an optical cavity is coupled to the motional \cite{Treutlein2013} or to the internal \cite{Vogell2015} degrees of freedom of a distant cloud of cold atoms that are trapped in the optical lattice of the outcoupled light field. By employing different cooling mechanisms, such as sympathetic cooling \cite{Treutlein2013,Treutlein2015,Zhong2017,Becker2018} and optical feedback cooling \cite{Becker2018,Treutlein2022,Treutlein2023}, the nanomembrane can be cooled down almost to its quantum ground state. Moreover, hybrid atom-optomechanical systems are a versatile playground for various quantum phenomena, such as indirect quantum measurement, atom-membrane entanglement and coherent state transfer \cite{Hammerer2009,Kimble2009,Wallquist2009,Palma2010,Genes2011}.

The internal state coupling scheme \cite{Vogell2015} is realized by a nanomembrane that is coupled to transitions between internal states of the atoms via translating the phase shift of the light, caused by the membrane displacement, into a polarization rotation using a polarizing beam splitter. This scheme allows for resonant coupling and mitigates the drawbacks of the motional coupling scheme \cite{Treutlein2013}, like strong frequency mismatch between the nanooscillator and the atomic motion in the optical trap. It has been utilized for membrane cooling \cite{Vogell2015,Lau2018}, displacement squeezing \cite{Ockeloen2013}, the realization of a effective negative mass oscillator \cite{Polzik2015}, and quantum back-action evading measurements \cite{Polzik2017}. In addition, a peculiar NQPT, whose order can be controlled by changing a directly accessible experimental parameter, has recently been proposed in this scheme \cite{Mann2019}. It has been found that the system can undergo both a first- and a second-order phase transition in the same physical setup, by tuning the atomic transition frequency.

In this work, we examine NQPTs in a hybrid atom-optomechanical system in the internal state coupling scheme. We consider an ultracold atomic gas trapped in a large quasi-1D box, where the lattice potentials generated by the light fields have been canceled. Since the membrane is naturally damped by its suspension, the atoms relax into a steady state whose features critically depend on the state of the membrane. For large enough trapping box and weak effective coupling between the atoms and the nanomembrane, we find that the atomic condensates are essentially uniform and the membrane is not displaced. When the atom-membrane couplings become strong enough, the system transitions into a symmetry-broken self-organized phase that exhibits density wave-like condensate profiles and significant nanomembrane displacement. Depending on the values of the readily controllable experimental parameters, this NQPT can be of either first or second order. We map out the relevant parts of the parameter space and study the hallmarks of the phase transition. Hence, the rest of the paper is organized as follows. In Sec.~\ref{sec:Theory} we outline the basic mean-field theory. In the following Sec.~\ref{sec:Hom} we examine the homogeneous steady states. Next, in Sec.~\ref{sec:Inhom} we focus on the inhomogeneous steady state and its features. We also address and examine the occurring 1st and 2nd order phase transitions. We discuss our results in Sec.~\ref{sec:Disscuss} and provide an outlook.

\section{Model and equations of motion}\label{sec:Theory}

\begin{figure}[h]
\centering
\includegraphics[width=\linewidth]{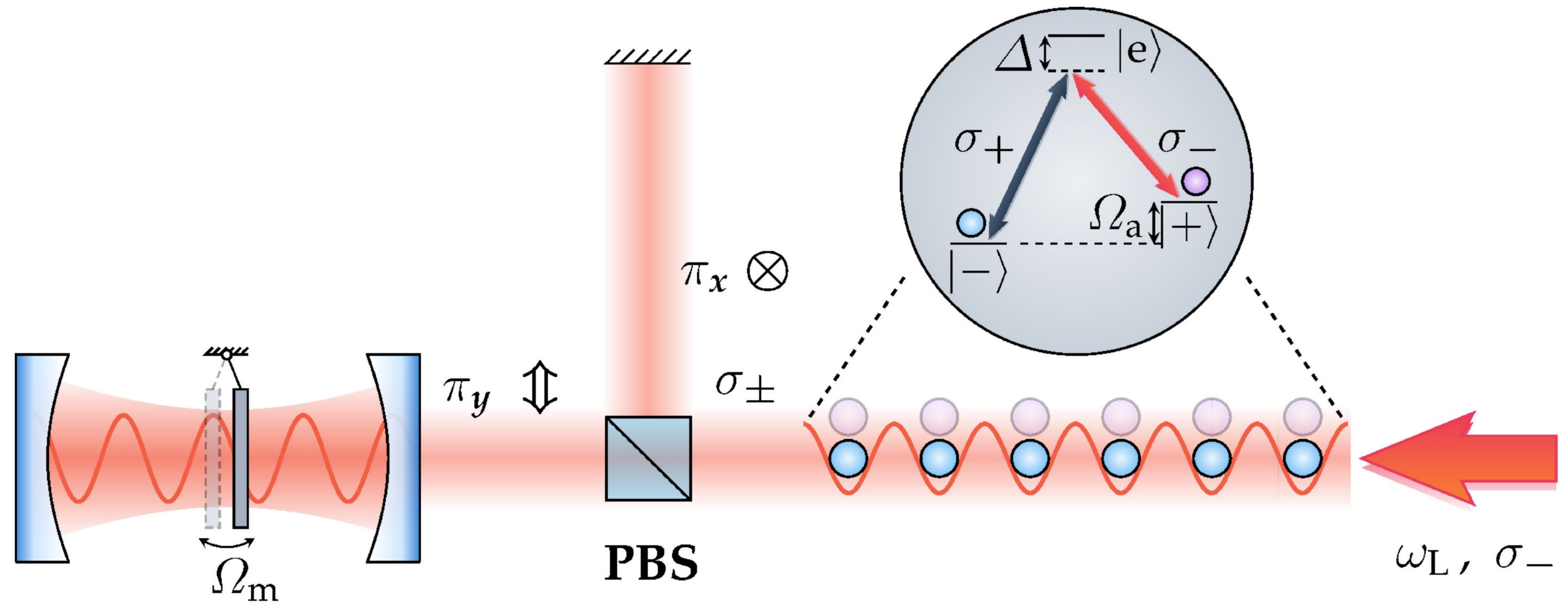}\vspace*{-3mm}
\caption{Semitransparent nanomembrane in an optical cavity is coupled to the internal states of a distant atomic ensemble that is trapped in an external quasi-1D trap. The internal states of the atoms constitute a $\Lambda$ level scheme according to the inset. The figure is adapted from Ref.\ \cite{Mann2019}.}
\label{fig:1}
\end{figure}

In the internal state coupling scheme \cite{Vogell2015}, the system of $N$ identical ultracold bosonic atoms is held in an external quasi-1D trap (see Fig.~\ref{fig:1}).
The relevant atomic states $\{|-\rangle, |+\rangle, |e\rangle \}$ constitute a $\Lambda$ level scheme, where the two lower states are energetically separated by the atomic transition frequency $\Omega_{\rm a}^{}$.
An applied $\sigma_-$ circularly polarized laser of frequency $\omega_{\rm L}^{}$ \tbl{and coherent field amplitude $\alpha_{\rm L}^{}$} drives the internal transition $|+\rangle\leftrightarrow |e\rangle$ at a finite large detuning $\Delta$.
The passing beam is sent to a polarizing beam splitter (PBS), which splits the circularly polarized light into linearly polarized $\pi_x$ and $\pi_y$ light beams in two perpendicular directions.
In the vertical arm, a fixed mirror simply reflects light with conserved
polarization $\pi_x$. The horizontal beam enters a low-finesse cavity which hosts a semitransparent nanomembrane with \tbl{the mass $M$ and} the mechanical resonance frequency $\Omega_{\rm m}^{}$.
The lengths of the two arms are made equal when the membrane is undisplaced. In that case, the cavity reflects $\pi_y$ light without any relative phase shift.
In a quasistatic picture, any finite displacement of the membrane induces a relative phase shift between the $\pi_x$ and $\pi_y$ light beams from the two arms.
This relative phase is converted to a polarization rotation after the two beams have passed backwards through the PBS.
The emerging $\sigma_+$ polarized light then drives the internal atomic transition $|-\rangle\leftrightarrow |e\rangle$ and may induce a two-photon Raman transition between the two lower states of the $\Lambda$ scheme via the excited state $|e\rangle$.
On the other hand, an internal transition between the two lower atomic states leads to the emission of a $\sigma_+$ polarized photon which alters the radiation pressure on the membrane after reaching it, which leads to an effective atom-membrane coupling.

\tbl{The Hamiltonian of this system has been derived in Ref.\ \cite{Vogell2015} and reduced to an effective form involving the membrane and the low-energy atomic degrees of freedom in Ref.\ \cite{Mann2019}. It reads (with $\hbar = c = 1$)}
\begin{widetext}\vspace*{-4mm}
\begin{align}\label{eq:H_eff}
\hat H_{\rm ef{}f}^{} &= \Omega_{\rm m}^{}\hat a^\dag\hat a + \sum_{\nu=\pm}\int dz\;\!\hat\psi_\nu^\dag(z)
\bigg[\nu\frac{\Omega_{\rm a}^{}}{2} - \omega_{\rm R}^{}\partial_z^2 + V_\nu^{}(z) + \frac{1}{2} \sum_{\nu'=\pm}g_{\nu\nu'}^{}\hat\psi_{\nu'}^\dag(z)\hat\psi_{\nu'}^{}(z)\bigg]\hat\psi_\nu^{}(z)\nonumber\\
& - (\hat a + \hat a^\dag)\int dz\:\!\sin(2z)\bigg[\lambda_{\rm ex}^{}\hat\psi_+^\dag(z)\hat\psi_+^{}(z) + \frac{\lambda}{2}\big(\hat\psi_+^\dag(z)\hat\psi_-^{}(z) + \hat\psi_-^\dag(z)\hat\psi_+^{}(z)\big)\bigg],
\end{align}
where the position coordinate has been re-scaled as $z\to z/\omega_{\rm L}^{}$. The bosonic operator $\hat{a}$ denotes the annihilation operator of the excitations of one single mechanical mode of the nanomembrane, while $\hat{\psi}_\nu^{}(z)$ denotes the field operator of the atoms in the state $\nu=\pm$. The atomic recoil frequency reads $\omega_{\rm R}^{} = \omega_{\rm L}^2/(2m)$, with $m$ being the atom mass, while $g_{\nu\nu'}^{}$ are atomic state-dependent contact interaction strengths. \tbl{The total potentials sensed by the atoms are denoted by $V_\pm^{}(z)$ and include the lattice potential for the atoms in the internal state $|+\rangle$ generated by the quadratic term in the coherent field amplitude $\alpha_{\rm L}^{}$ \cite{Mann2019}, as well as the external potentials. The interaction of the light field with the atoms also has a linear effect in $\alpha_{\rm L}^{}$ that manifests itself in two essentially different ways, depending on the atomic state. The first drives the internal transition $|+\rangle\leftrightarrow |e\rangle$ via the $\sigma_-$ field quadrature in a manner similar to the motional coupling scheme. The second causes Raman transitions $|+\rangle\leftrightarrow |-\rangle$ under the creation (or annihilation) of a $\sigma_+$ polarized photon. After adiabatic elimination of the excited state $|e\rangle$ and the light field one obtains the effective membrane-atom interaction, described by the two terms in the second line of the Hamiltonian \eqref{eq:H_eff}, with the effective coupling strengths $\lambda_{\rm ex}^{}$ and $\lambda$, respectively, which are quadratic in $\alpha_{\rm L}^{}$ and can be expressed in terms of the physical parameters of the full model as $\lambda_{\rm ex}^{}= \lambda_{\rm m}\lambda_{\rm a}/2$ and $\lambda=\lambda_{\rm m}\lambda_{\pm}/2$ \cite{Mann2019}. The light-membrane coupling constant is given by $\lambda_{\rm m}=2 \alpha_{\rm L} |r_{\rm m}| \omega_{\rm L} \ell_{\rm m} {\cal F}/\pi^{3/2}$, with $|r_{\rm m}|$ being the membrane reflectivity, $\ell_{\rm m}=(1/M\Omega_{\rm m})^{-1/2}$ the amplitude of the zero-point motion of the membrane, and ${\cal F}$ the cavity finesse. Moreover, the atomic intraspecies coupling constant is given by $\lambda_{\rm a}=\sqrt{2\pi} \alpha_{\rm L} \omega_{\rm L} \mu_+^2 {\cal E}_{\omega_{\rm L}}/\Delta$, where $\mu_+$ is the atomic dipole moment of the transition $|+\rangle\leftrightarrow |e\rangle$ and ${\cal E}_{\omega_{\rm L}}=\sqrt{\omega_{\rm L}/\pi {\cal A}}$ is a normalization constant of the light mode operators with ${\cal A}$ being the laser beam cross-sectional area. $\Delta$ is the (large) detuning between the frequency of the pump laser and the addressed internal atomic transition. Furthermore, the atomic interspecies coupling constant is given by $\lambda_{\pm} =\sqrt{2\pi} \alpha_{\rm L} \omega_{\rm L} \mu_+ \mu_- {\cal E}_{\omega_{\rm L}}/\Delta$, where $\mu_-$ is the atomic dipole moment of the transition $|-\rangle\leftrightarrow |e\rangle$.}
 \tbl{Note that the effective cavity-mediated global atom-atom interaction has been neglected, since our present focus is on the parameter regime where it is weak enough not to affect the system periodicity \cite{Zhou2021}.} Taking into account that the membrane is damped at the rate $\Gamma_{\rm m}^{}$, we derive the following equations of motion for the membrane and the atomic operators
\begin{subequations}\label{eq:OP}
\begin{align}
i\partial_t^{}\hat a &= \big(\Omega_{\rm m}^{}-i\Gamma_{\rm m}^{}\big)\hat a - \int dz\:\!\sin(2z) \bigg[\lambda_{\rm ex}^{}\hat\psi_+^\dag(z)\hat\psi_+^{}(z) + \frac{\lambda}{2}\big(\hat\psi_+^\dag(z)\hat\psi_-^{}(z) + \hat\psi_-^\dag(z)\hat\psi_+^{}(z)\big)\bigg] - \hat\xi_{\rm m}^{},\displaybreak[0]\\
i\partial_t^{}\hat\psi_-^{}(z) &= \bigg[-\frac{\Omega_{\rm a}^{}}{2} - \omega_{\rm R}^{}\partial_z^2 + V_-^{}(z) + \sum_{\nu=\pm}g_{-\nu}^{}\hat\psi_\nu^\dag(z)\hat\psi_\nu^{}(z)\bigg]\hat\psi_-^{}(z) - \frac{\lambda}{2}\sin(2z)\big(\hat a + \hat a^\dag\big)\hat\psi_+^{}(z),\displaybreak[0]\\
i\partial_t^{}\hat\psi_+^{}(z) &= \bigg[\frac{\Omega_{\rm a}^{}}{2} - \omega_{\rm R}^{}\partial_z^2 + V_+^{}(z) + \sum_{\nu=\pm}g_{+\nu}^{}\hat\psi_\nu^\dag(z)\hat\psi_\nu^{}(z)\bigg]\hat\psi_+^{}(z) - \frac{\lambda}{2}\sin(2z)\big(\hat a + \hat a^\dag\big)\hat\psi_-^{}(z)\nonumber\\
& - \lambda_{\rm ex}^{}\sin(2z)\big(\hat a + \hat a^\dag\big)\hat\psi_+^{}(z),
\end{align}
\end{subequations}
\end{widetext}
where we have omitted the time argument for brevity. In accordance with the damping of the membrane mode, we have introduced the corresponding bosonic noise operator $\hat\xi_{\rm m}^{}$ that is characterized by the zero mean fluctuations $\avg{\hat\xi_{\rm m}^{}}=\avg{\hat\xi_{\rm m}^\dag}=0$, as well as by the auto-correlations
\begin{subequations}\vspace*{-0.6mm}
\begin{align}
\avg{\hat\xi_{\rm m}^{}(t)\hat\xi_{\rm m}^\dag(0)} &= 2\Gamma_{\rm m}^{}(n_{\rm m}^{} + 1)\delta(t),\displaybreak[0]\\
\avg{\hat\xi_{\rm m}^\dag(t)\hat\xi_{\rm m}^{}(0)} &= 2\Gamma_{\rm m}^{}n_{\rm m}^{}\delta(t).
\end{align}
\end{subequations}
The environment occupation number $n_{\rm m}^{}$ determines the steady-state occupation of the membrane mode when it is affected only by its environment \cite{Zoller2004}.

\tbl{We assume that the atoms are at ultra-low temperatures and that the atom-membrane coupling is such that a large fraction of the atoms condense and the system dynamics can be well described in first approximation at the mean-field level. The influence of quantum fluctuations is an interesting topic in itself and will be addressed in the follow-up publication. Here we focus on the coupled mean-field equations of motion}
\begin{widetext}\vspace*{-4mm}
\begin{subequations}\label{eq:MF}
\begin{align}
i\partial_t^{}\alpha &= \big(\Omega_{\rm m}^{}-i\Gamma_{\rm m}^{}\big)\alpha - \int dz\:\!\sin(2z) \bigg[\sqrt{N}\lambda_{\rm ex}^{} |\psi_+^{}(z)|^2 + \frac{\sqrt{N}\lambda}{2}\big(\psi_+^*(z)\psi_-^{}(z) + \psi_-^*(z)\psi_+^{}(z)\big)\bigg],\displaybreak[0]\\
i\partial_t^{}\psi_-^{}(z) &= \bigg[-\frac{\Omega_{\rm a}^{}}{2} - \omega_{\rm R}^{}\partial_z^2 + V_-^{}(z) + N\sum_{\nu=\pm}g_{-\nu}^{}|\psi_\nu^{}(z)|^2\bigg]\psi_-^{}(z) - \frac{\sqrt{N}\lambda}{2}\sin(2z)\big(\alpha + \alpha^*\big)\psi_+^{}(z),\displaybreak[0]\\
i\partial_t^{}\psi_+^{}(z) &= \bigg[\frac{\Omega_{\rm a}^{}}{2} - \omega_{\rm R}^{}\partial_z^2 + V_+^{}(z) + N\sum_{\nu=\pm}g_{+\nu}^{}|\psi_\nu^{}(z)|^2\bigg]\psi_+^{}(z) - \frac{\sqrt{N}\lambda}{2}\sin(2z)\big(\alpha + \alpha^*\big)\psi_-^{}(z)\nonumber\displaybreak[0]\\
& - \sqrt{N}\lambda_{\rm ex}^{} \sin(2z)\big(\alpha + \alpha^*\big)\psi_+^{}(z),
\end{align}
\end{subequations}
where we have neglected all mutual correlations and introduced the complex coherent state amplitude of the membrane $\alpha = \avg{\hat a}/\sqrt{N}$, and the wave functions of the two condensates $\psi_\pm^{}(z) = \avg{\hat\psi_\pm^{}(z)}/\sqrt{N}$. We assume that the membrane damping is so fast that the membrane almost instantaneously relaxes to its steady state determined by the condensate wave functions
\begin{align}
\bar\alpha &= \frac{1}{\Omega_{\rm m}^{}-i\Gamma_{\rm m}^{}}\int dz\:\!\sin(2z) \bigg[\sqrt{N}\lambda_{\rm ex}^{} |\psi_+^{}(z)|^2 + \frac{\sqrt{N}\lambda}{2}\big(\psi_+^*(z)\psi_-^{}(z) + \psi_-^*(z)\psi_+^{}(z)\big)\bigg],
\end{align}
so that we obtain the following coupled equations for the two condensates
\begin{subequations}\label{eq:MF-cond}
\begin{align}
i\partial_t^{}\psi_-^{}(z) &= \bigg[-\frac{\Omega_{\rm a}^{}}{2} - \omega_{\rm R}^{}\partial_z^2 + V_-^{}(z) + N\sum_{\nu=\pm}g_{-\nu}^{}|\psi_\nu^{}(z)|^2\bigg]\psi_-^{}(z) - \frac{\sqrt{N}\lambda}{2}\sin(2z)\big(\bar\alpha + \bar\alpha^*\big)\psi_+^{}(z),\displaybreak[0]\\
i\partial_t^{}\psi_+^{}(z) &= \bigg[\frac{\Omega_{\rm a}^{}}{2} - \omega_{\rm R}^{}\partial_z^2 + V_+^{}(z) + N\sum_{\nu=\pm}g_{+\nu}^{}|\psi_\nu^{}(z)|^2\bigg]\psi_+^{}(z) - \frac{\sqrt{N}\lambda}{2}\sin(2z)\big(\bar\alpha + \bar\alpha^*\big)\psi_-^{}(z)\nonumber\displaybreak[0]\\
& - \sqrt{N}\lambda_{\rm ex}^{}\sin(2z)\big(\bar\alpha + \bar\alpha^*\big)\psi_+^{}(z).
\end{align}
\end{subequations}
Due to the damping of the membrane, the atoms also eventually reach the minimal-energy steady state, i.e., the ground steady state (GSS), described by the ansatz $\psi_\pm^{}(z,t)=e^{-i\mu t}\psi_\pm^{}(z)$. It involves the chemical potential $\mu$ that is fixed by the normalization condition
\begin{align}\label{eq:norm_cond}
\int dz\;\!\big(|\psi_-^{}(z)|^2 + |\psi_+^{}(z)|^2\big) = 1.
\end{align}
It follows that the GSS solutions of the equations of motion \eqref{eq:MF-cond} can be obtained either by solving the stationary equations
\begin{subequations}\label{eq:MF-cond-stat}
\begin{align}
0 &= \bigg[-\frac{\Omega_{\rm a}^{}}{2} - \omega_{\rm R}^{}\partial_z^2 - \mu + V_-^{}(z) + N\sum_{\nu=\pm}g_{-\nu}^{}|\psi_\nu^{}(z)|^2\bigg]\psi_-^{}(z) - \frac{\sqrt{N}\lambda}{2}\sin(2z)\big(\bar\alpha + \bar\alpha^*\big)\psi_+^{}(z),\displaybreak[0]\\
0 &= \bigg[\frac{\Omega_{\rm a}^{}}{2} - \omega_{\rm R}^{}\partial_z^2- \mu + V_+^{}(z) + N\sum_{\nu=\pm}g_{+\nu}^{}|\psi_\nu^{}(z)|^2\bigg]\psi_+^{}(z) - \frac{\sqrt{N}\lambda}{2}\sin(2z)\big(\bar\alpha + \bar\alpha^*\big)\psi_-^{}(z)\nonumber\displaybreak[0]\\
& - \sqrt{N}\lambda_{\rm ex}^{}\sin(2z)\big(\bar\alpha + \bar\alpha^*\big)\psi_+^{}(z),
\end{align}
\end{subequations}
or, equivalently, by minimizing the energy functional
\begin{align}\label{eq:E}
\mc{E}_N^{}[\psi_\pm^{}] &= N\sum_{\nu=\pm}\int dz\;\!\psi_\nu^*(z)
\bigg[\nu\frac{\Omega_{\rm a}^{}}{2} - \omega_{\rm R}^{}\partial_z^2 + V_\nu^{}(z) + \frac{N}{2}\sum_{\nu'=\pm}g_{\nu\nu'}^{}|\psi_{\nu'}^{}(z)|^2\bigg] \psi_\nu^{}(z)\nonumber\\
& - \frac{N^2\Omega_{\rm m}^{}}{\Omega_{\rm m}^2 + \Gamma_{\rm m}^2}\bigg\{\int dz\:\!\sin(2z)\bigg[\lambda_{\rm ex}^{}|\psi_+^{}(z)|^2 + \frac{\lambda}{2}\big(\psi_+^*(z)\psi_-^{}(z) + \psi_-^*(z)\psi_+^{}(z)\big)\bigg]\bigg\}^2,
\end{align}
with respect to $\psi_\pm^*(z)$ and under the normalization constraint \eqref{eq:norm_cond}. This minimization justifies our notion of the GSS. The chemical potential stems from the relation $\mu=\partial\mc{E}_N^{}[\psi_\pm^{}]/\partial N$ and reads
\begin{align}\label{eq:mu}
\mu &= \sum_{\nu=\pm}\int dz\;\!\psi_\nu^*(z)
\bigg[\nu\frac{\Omega_{\rm a}^{}}{2} - \omega_{\rm R}^{}\partial_z^2 + V_\nu^{}(z) + N\sum_{\nu'=\pm}g_{\nu\nu'}^{}|\psi_{\nu'}^{}(z)|^2\bigg] \psi_\nu^{}(z)\nonumber\\
& - \frac{2N\Omega_{\rm m}^{}}{\Omega_{\rm m}^2 + \Gamma_{\rm m}^2}\bigg\{\int dz\:\!\sin(2z)\bigg[\lambda_{\rm ex}^{}|\psi_+^{}(z)|^2 + \frac{\lambda}{2}\big(\psi_+^*(z)\psi_-^{}(z) + \psi_-^*(z)\psi_+^{}(z)\big)\bigg]\bigg\}^2.
\end{align}

Let us next examine the ground steady state of the system. \tbl{The external trapping potentials are designed to {\it cancel} the lattice potentials for the atoms in the internal state $|+\rangle$, generated by the quadratic term in the coherent field amplitude \cite{Mann2019}, and lead to the box-like traps $V_\pm^{}(z)$ with the hard walls at $z\in\{z_1^{},z_2^{}\}$, where $z_1^{}=-\pi/4-\ell\pi$, $z_2^{}=-\pi/4+\ell\pi$ for some large integer $\ell\gtrsim 100$.} That imposes the Dirichlet boundary conditions $\psi_\pm^{}(z_1^{})=\psi_\pm^{}(z_2^{})=0$. According to our assumption $\ell\gg 1$, we can safely neglect any edge effects that significantly affect the condensates only within a few healing lengths of the walls. Since the atom-membrane interaction is periodic with the period of $\pi$, we can reduce our attention onto the single period $[-\pi/4,3\pi/4]$. We impose the periodic boundary conditions $\psi_-^{}(-\pi/4)=\psi_-^{}(3\pi/4)$, $\psi_+^{}(-\pi/4)=\psi_+^{}(3\pi/4)$ and take $V_\pm^{}(z)=0$ henceforward. Note that the spatial dependence of the atom-membrane interaction, $\propto \sin(2z)$, is even with respect to the center of the chosen period. This brings us to the energy per period
\begin{align}\label{eq:E_pi}
\tilde{\mc{E}}_{\tilde{N}}^{}[\tilde{\psi}_\pm^{}] &\equiv \frac{\mc{E}_{N}^{}[\psi_\pm^{}]}{2\ell} = \tilde{N}\sum_{\nu=\pm}\int_\pi dz\;\!\tilde{\psi}_\nu^*(z)
\bigg[\nu\frac{\Omega_{\rm a}^{}}{2} - \omega_{\rm R}^{}\partial_z^2 + \frac{\tilde{N}}{2}\sum_{\nu'=\pm}g_{\nu\nu'}^{}|\tilde{\psi}_{\nu'}^{}(z)|^2\bigg]\tilde{\psi}_\nu^{}(z)\nonumber\\
& - \frac{\tilde{N}^2\Omega_{\rm m}^{}}{\Omega_{\rm m}^2 + \Gamma_{\rm m}^2}\bigg\{\int_\pi dz\:\!\sin(2z)\bigg[\tilde{\lambda}_{\rm ex}^{}|\tilde{\psi}_+^{}(z)|^2 + \frac{\tilde{\lambda}}{2}\big(\tilde{\psi}_+^*(z)\tilde{\psi}_-^{}(z) + \tilde{\psi}_-^*(z)\tilde{\psi}_+^{}(z)\big)\bigg]\bigg\}^2,
\end{align}
where we introduced the rescaled quantities
\begin{align}\label{eq:rescale}
N = 2\ell\tilde{N},\quad\psi_\pm^{}(z) = \tilde{\psi}_\pm^{}(z)/\sqrt{2\ell},
\quad\lambda = \tilde{\lambda}/\sqrt{2\ell},
\quad\lambda_{\rm ex}^{} = \tilde{\lambda}_{\rm ex}^{}/\sqrt{2\ell}.
\end{align}
The integrals in Eq.~(\ref{eq:E_pi}) run over one single period of length $\pi$. In terms of the above, the membrane amplitude becomes
\begin{align}
\bar\alpha &= \frac{1}{\Omega_{\rm m}^{}-i\Gamma_{\rm m}^{}}\int_\pi dz\:\!\sin(2z) \bigg[\sqrt{\tilde{N}}\tilde{\lambda}_{\rm ex}^{}|\tilde{\psi}_+^{}(z)|^2 + \frac{\sqrt{\tilde{N}}\tilde{\lambda}}{2}\big(\tilde{\psi}_+^*(z)\tilde{\psi}_-^{}(z) + \tilde{\psi}_-^*(z)\tilde{\psi}_+^{}(z)\big)\bigg],
\end{align}
while the rescaled wave functions evolve in accordance with
\begin{subequations}\label{eq:MF-cond_pi}
\begin{align}
i\partial_t^{}\tilde{\psi}_-^{}(z) &= \bigg[-\frac{\Omega_{\rm a}^{}}{2} - \omega_{\rm R}^{}\partial_z^2 + \tilde{N}\sum_{\nu=\pm}g_{-\nu}^{}|\tilde{\psi}_\nu^{}(z)|^2\bigg]\tilde{\psi}_-^{}(z) - \frac{\sqrt{\tilde{N}}\tilde{\lambda}}{2}\sin(2z)\big(\bar\alpha + \bar\alpha^*\big)\tilde{\psi}_+^{}(z),\displaybreak[0]\\
i\partial_t^{}\tilde{\psi}_+^{}(z) &= \bigg[\frac{\Omega_{\rm a}^{}}{2} - \omega_{\rm R}^{}\partial_z^2 + \tilde{N}\sum_{\nu=\pm}g_{+\nu}^{}|\tilde{\psi}_\nu^{}(z)|^2\bigg]\tilde{\psi}_+^{}(z) - \frac{\sqrt{\tilde{N}}\tilde{\lambda}}{2}\sin(2z)\big(\bar\alpha + \bar\alpha^*\big)\tilde{\psi}_-^{}(z)\nonumber\displaybreak[0]\\
& - \sqrt{\tilde{N}}\tilde{\lambda}_{\rm ex}^{}\sin(2z)\big(\bar\alpha + \bar\alpha^*\big)\tilde{\psi}_+^{}(z),
\end{align}
\end{subequations}
and are normalized according to
\begin{align}
\int_\pi dz\;\!\big(|\tilde{\psi}_-^{}(z)|^2 + |\tilde{\psi}_+^{}(z)|^2\big) = 1.
\end{align}
The stationary wave function equations, which determine the GSS, turn into
\begin{subequations}\label{eq:MF-cond-stat_pi}
\begin{align}
0 &= \bigg[-\frac{\Omega_{\rm a}^{}}{2} - \omega_{\rm R}^{}\partial_z^2 - \mu + \tilde{N}\sum_{\nu=\pm}g_{-\nu}^{}|\tilde{\psi}_\nu^{}(z)|^2\bigg]\tilde{\psi}_-^{}(z) - \frac{\sqrt{\tilde{N}}\tilde{\lambda}}{2}\sin(2z)\big(\bar\alpha + \bar\alpha^*\big)\tilde{\psi}_+^{}(z),\displaybreak[0]\\
0 &= \bigg[\frac{\Omega_{\rm a}^{}}{2} - \omega_{\rm R}^{}\partial_z^2 - \mu + \tilde{N}\sum_{\nu=\pm}g_{+\nu}^{}|\tilde{\psi}_\nu^{}(z)|^2\bigg]\tilde{\psi}_+^{}(z) - \frac{\sqrt{\tilde{N}}\tilde{\lambda}}{2}\sin(2z)\big(\bar\alpha + \bar\alpha^*\big)\tilde{\psi}_-^{}(z)\nonumber\displaybreak[0]\\
& - \sqrt{\tilde{N}}\tilde{\lambda}_{\rm ex}^{}\sin(2z)\big(\bar\alpha + \bar\alpha^*\big)\tilde{\psi}_+^{}(z),
\end{align}
\end{subequations}
where the chemical potential is given by
\begin{align}\label{eq:mu_pi}
\mu &= \sum_{\nu=\pm}\int_\pi dz\;\!\tilde{\psi}_\nu^*(z)
\bigg[\nu\frac{\Omega_{\rm a}^{}}{2} - \omega_{\rm R}^{}\partial_z^2 + \tilde{N}\sum_{\nu'=\pm}g_{\nu\nu'}^{}|\tilde{\psi}_{\nu'}^{}(z)|^2\bigg] \tilde{\psi}_\nu^{}(z)\nonumber\\
& - \frac{2\tilde{N}\Omega_{\rm m}^{}}{\Omega_{\rm m}^2 + \Gamma_{\rm m}^2}\bigg\{\int_\pi dz\:\!\sin(2z)\bigg[\tilde{\lambda}_{\rm ex}^{}|\tilde{\psi}_+^{}(z)|^2 + \frac{\tilde{\lambda}}{2}\big(\tilde{\psi}_+^*(z)\tilde{\psi}_-^{}(z) + \tilde{\psi}_-^*(z)\tilde{\psi}_+^{}(z)\big)\bigg]\bigg\}^2.
\end{align}
\end{widetext}

In the following, we will assume that the rescalings \eqref{eq:rescale} have been performed, and, for the sake of brevity, we will omit the tildes. All extensive quantities addressed below are therefore per period.

\section{Homogeneous ground steady states}\label{sec:Hom}

We first analyze the homogeneous solutions using the ansatz $\psi_\pm^{}(z)=\sqrt{n_\pm^{}/\pi}$, where $n_\pm^{}$ are the occupation fractions of the two atomic states. In such a case, the normalization condition simply reads $n_-^{} + n_+^{} = 1$ and the membrane is not excited, i.e., $\bar\alpha=0$. The stationary equations \eqref{eq:MF-cond-stat_pi} become
\begin{subequations}\label{eq:MF-hom}
\begin{align}
0 &= \bigg[-\frac{\Omega_{\rm a}^{}}{2} - \mu + \frac{N}{\pi}\sum_{\nu=\pm}g_{-\nu}^{}n_\nu^{}\bigg] \sqrt{n_-^{\vphantom{*}}}\:\!,\displaybreak[0]\\
0 &= \bigg[\frac{\Omega_{\rm a}^{}}{2} - \mu + \frac{N}{\pi}\sum_{\nu=\pm}g_{+\nu}^{}n_\nu^{}\bigg] \sqrt{n_+^{\vphantom{*}}}\:\!.
\end{align}
\end{subequations}
We will be considering the on average repulsive regime where $g_{--}^{},g_{++}^{}>0>g_{+-}^{}$ and $g_{--}^{}g_{++}^{}>g_{+-}^2$. Three solutions are possible, which read
\begin{subequations}
\label{eq:hom-sols}
\begin{eqnarray}
n_-^{} &=& 1,\quad n_+^{} = 0,\quad \mu = -\frac{\Omega_{\rm a}^{}}{2} + \frac{N}{\pi} g_{--}^{},\\
n_-^{} &=& 0,\quad n_+^{} = 1,\quad \mu = \frac{\Omega_{\rm a}^{}}{2} + \frac{N}{\pi} g_{++}^{},
\end{eqnarray}
\text{as well as}
\begin{align}
n_-^{} ={}& \frac{g_{++}^{}-g_{+-}^{}+\pi\Omega_{\rm a}^{}/N}{g_{--}^{}+g_{++}^{}-2g_{+-}^{}},\displaybreak[0]\nonumber\\
n_+^{} ={}& \frac{g_{--}^{}-g_{+-}^{}-\pi\Omega_{\rm a}^{}/N}{g_{--}^{}+g_{++}^{}-2g_{+-}^{}},\displaybreak[0]\nonumber\\
\mu ={}& \frac{\Omega_{\rm a}^{}}{2}\frac{g_{--}^{}-g_{++}^{}}{g_{--}^{}+g_{++}^{}-2g_{+-}^{}}\nonumber\\
& + \frac{N}{\pi}\frac{g_{--}^{}g_{++}^{}-g_{+-}^2}{g_{--}^{}+g_{++}^{}-2g_{+-}^{}}.
\end{align}
\end{subequations}
The first solution is the lowest in energy when $\pi\Omega_{\rm a}^{}/N>g_{--}^{}-g_{+-}^{}$ holds, while for $\pi\Omega_{\rm a}^{}/N<g_{+-}^{}-g_{++}^{}$, the second one takes over. The third solution exists only if $g_{+-}^{}-g_{++}^{}<\pi\Omega_{\rm a}^{}/N<g_{--}^{}-g_{+-}^{}$ is satisfied and has the lowest energy of the three. For $\pi\Omega_{\rm a}^{}/N<(g_{--}^{}-g_{++}^{})/2$ we have $n_-^{}<n_+^{}$, while for $\pi\Omega_{\rm a}^{}/N\ge (g_{--}^{}-g_{++}^{})/2$ we get $n_-^{}\ge n_+^{}$. Thus, it is convenient to introduce the reference transition frequency $\Omega_{\rm a,0}^{} = N(g_{--}^{}-g_{++}^{})/(2\pi)$ and study the behavior of the system as a function of the {\it relative detuning}
\begin{align}
\delta\Omega_{\rm a}^{}\equiv\Omega_{\rm a}^{}-\Omega_{\rm a,0}^{},
\end{align}
of the atomic transition frequency from it. For strong enough couplings $\lambda$ and $\lambda_{\rm ex}^{}$ the system transitions into a self-organized GSS. The membrane becomes excited and the condensates become inhomogeneous. The nature of this phase transition is what we are going to investigate next.

\section{Phase transitions to and from inhomogeneous ground steady state}\label{sec:Inhom}

Let us now explore the self-organized phase. To this end, we evolve the equations of motion \eqref{eq:MF-cond_pi} in imaginary time starting with a suitable initial state until the ground steady state is reached. As can be seen from these equations, the membrane couples to the condensates via the combination ${\rm Re}[\bar\alpha]\sin(2z)$. \tbl{For positive ${\rm Re}[\bar\alpha]$, the inhomogeneous GSS wave functions must reach their maximum together with $\sin(2z)$, i.e., in the middle of the chosen period. In the opposite case of the negative real part, the wave functions are simply shifted by half a period. Therefore, we will only consider the positive case.} Also, $\psi_\pm^{}(z)$ have to be even functions with respect to the same point, in the same manner as $\sin(2z)$. In this way, the atom-membrane interaction term in Eq.\ \eqref{eq:E_pi} contributes most to the minimization of the energy of the system.
\begin{figure}[b]
\centering
\includegraphics[width=0.9\linewidth]{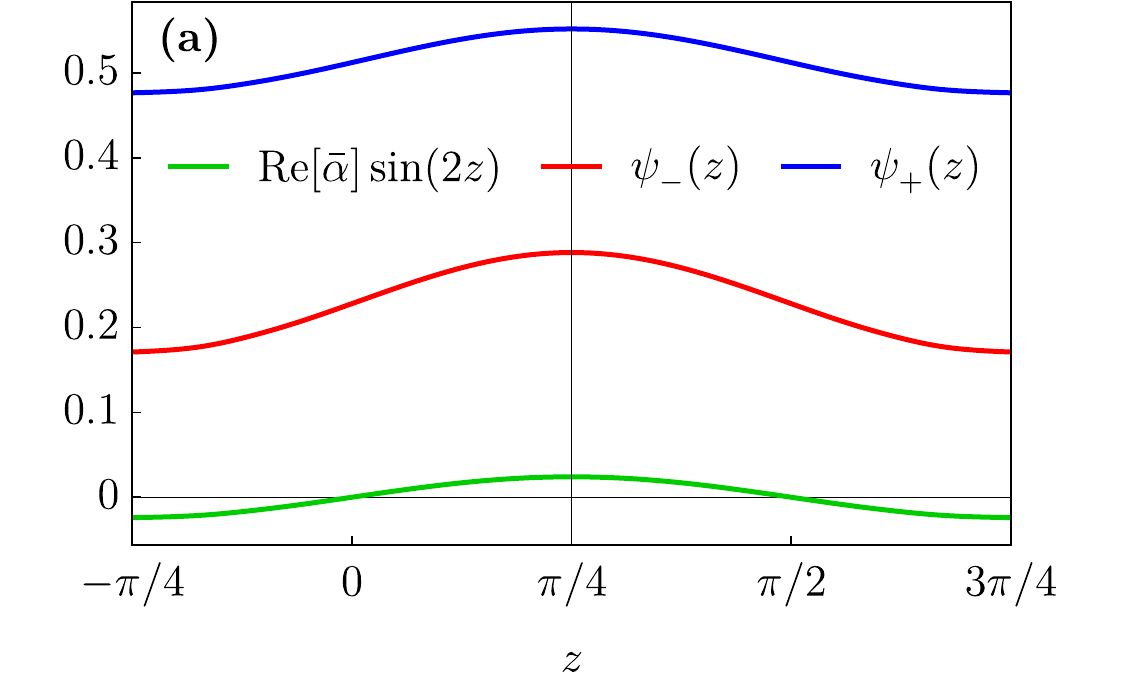}
\\[3mm]
\hspace*{-2.5mm}
\includegraphics[width=0.833\linewidth]{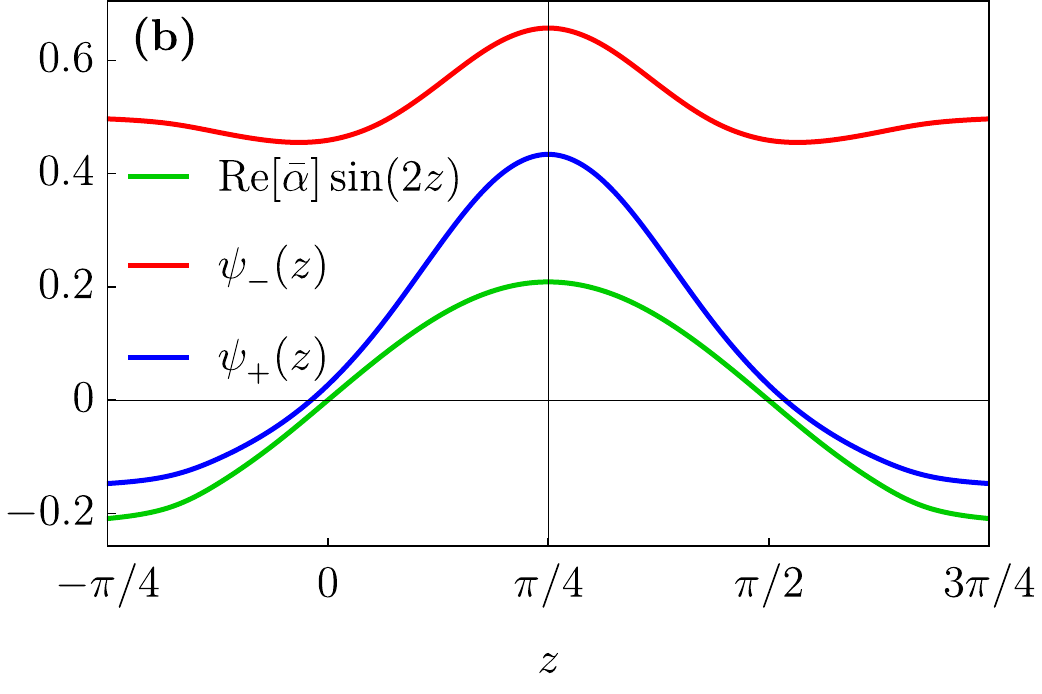}
\vspace*{-2mm}
\caption{Spatial profiles of ${\rm Re}[\bar\alpha]\sin(2z)$ (green) and the ground steady state condensate wave functions $\psi_-^{}(z)$ (red) and $\psi_+^{}(z)$ (blue) over one period for large negative (a) and large positive (b) relative detuning. Parameters: $\Omega_{\rm m}^{} = 100\omega_{\rm R}^{}$, $\Gamma_{\rm m}^{} = 10\omega_{\rm R}^{}$, $N g_{--}^{} = N g_{++}^{} = 100\omega_{\rm R}^{}$, $N g_{+-}^{} = -90\omega_{\rm R}^{}$, and $\sqrt{N}\lambda_{\rm ex}^{} = 35\omega_{\rm R}^{}$, while in (a) $\delta\Omega_{\rm a}^{} = -40\omega_{\rm R}^{}$, $\sqrt{N}\lambda = 5\omega_{\rm R}^{}$ and in (b) $\delta\Omega_{\rm a}^{} = 70\omega_{\rm R}^{}$, $\sqrt{N}\lambda = 70\omega_{\rm R}^{}$.}
\label{fig:2}
\end{figure}
Thus, the combination ${\rm Re}[\bar\alpha]\sin(2z)$ captures the essence of the membrane's influence on the atoms, both in terms of its spatial behavior and its relative strength. Figure \ref{fig:2} shows two exemplary cases of spatial profiles of ${\rm Re}[\bar\alpha]\sin(2z)$, together with the two GSS wave functions $\psi_\pm^{}(z)$ over the chosen period. In panel (a) we present the case of a negative relative detuning $\delta\Omega_{\rm a}^{}$, where the $|+\rangle$ state usually has higher occupation. Panel (b) displays the opposite case with the $|-\rangle$ state being more populated. As can be seen, larger spatial variances of the wave functions correspond to larger membrane amplitude.
\begin{figure}
\centering
\includegraphics[width=0.83\linewidth]{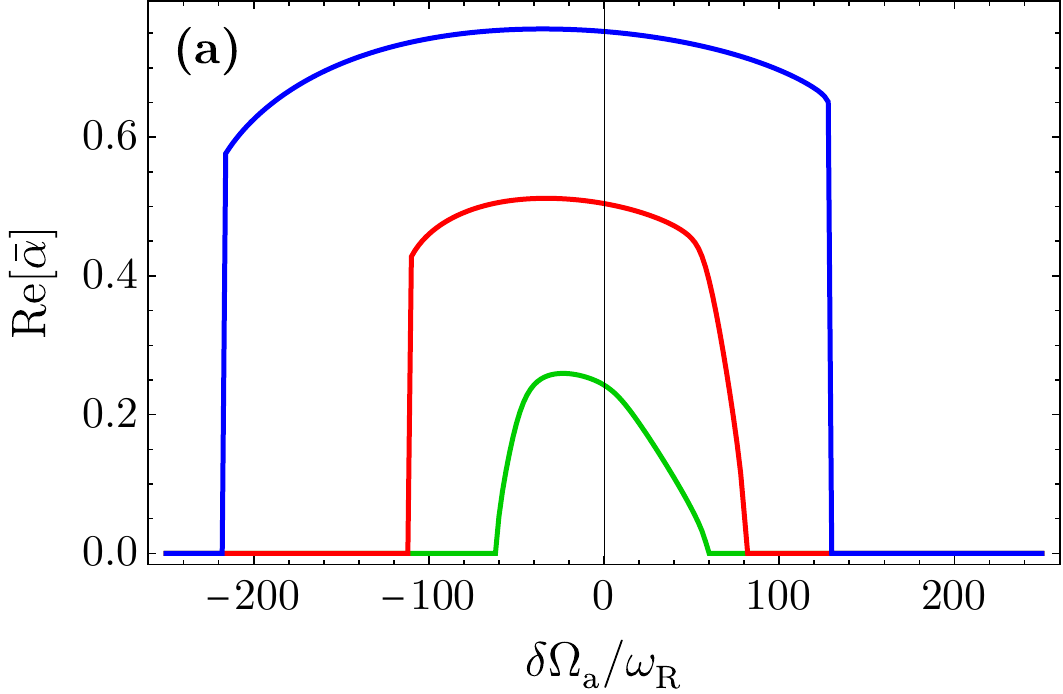}
\\[3mm]
\includegraphics[width=0.845\linewidth]{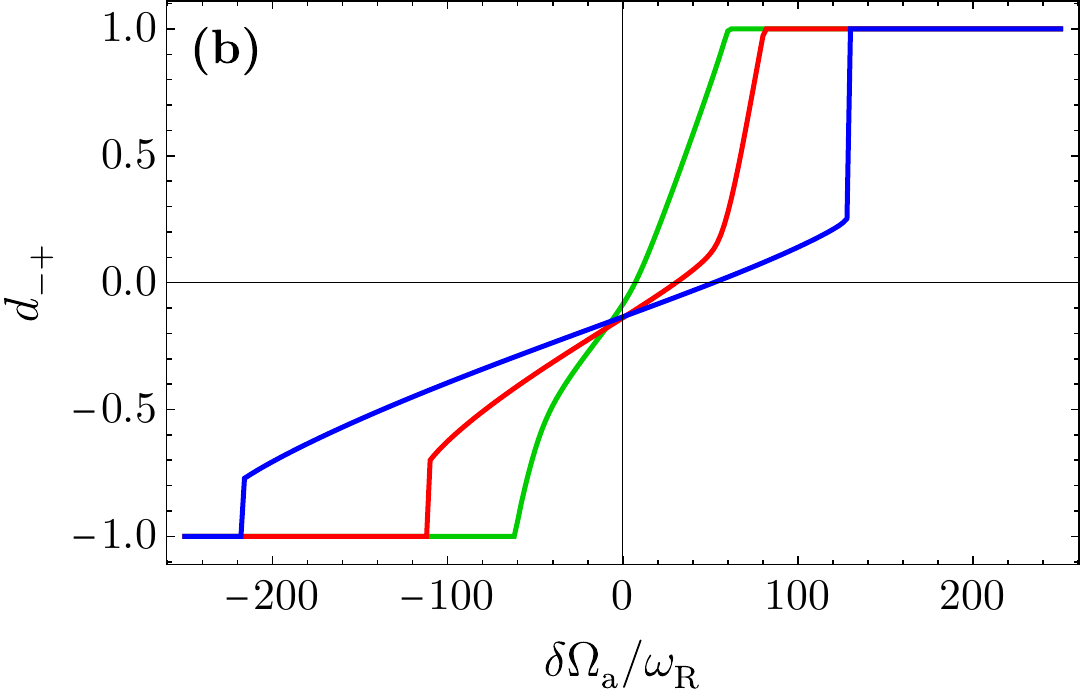}
\vspace*{-2mm}
\caption{Dependence of the real part of the membrane amplitude ${\rm Re}[\bar\alpha]$ (a) and the population imbalance $d_{-+}^{}\equiv(N_-^{}-N_+^{})/N$ (b) between the $|-\rangle$ and the $|+\rangle$ state on the relative detuning $\delta\Omega_{\rm a}^{}$ for a fixed value of the coupling $\sqrt{N}\lambda_{\rm ex}^{}=40\omega_{\rm R}^{}$ and for three values of the coupling $\sqrt{N}\lambda\in\{20,70,120\}\;\!\omega_{\rm R}^{}$ (green, red, blue curves). Other parameters: $\Omega_{\rm m}^{} = 100\omega_{\rm R}^{}$, $\Gamma_{\rm m}^{} = 10\omega_{\rm R}^{}$, $N g_{--}^{} = N g_{++}^{} = 100\omega_{\rm R}^{}$, and $N g_{+-}^{} = -90\omega_{\rm R}^{}$.}
\label{fig:3}
\end{figure}
For large negative relative detunings, the condensate is homogeneous and the atoms occupy exclusively the $|+\rangle$ state. Likewise, for large positive $\delta\Omega_{\rm a}^{}$ the atoms homogeneously condense into the $|-\rangle$ state. In between, both atomic states exhibit a macroscopic population, while the system may transition from the homogeneous to the self-organized phase. Such transitions occur for sufficiently large couplings $\lambda$ and $\lambda_{\rm ex}^{}$ to the membrane. Their values determine whether the transitions are continuous or discontinuous.

\begin{figure*}
\centering
\hspace*{-1mm}\includegraphics[width=0.747\linewidth]{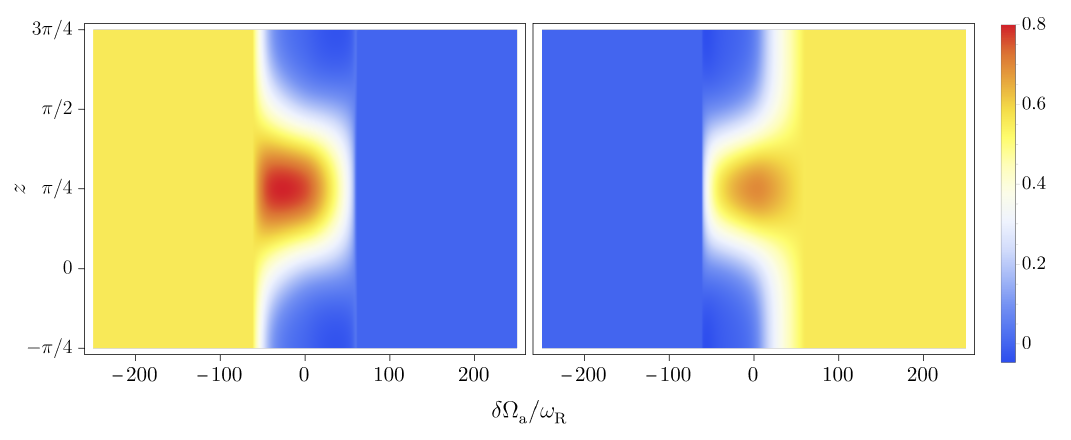}\\
\hspace*{-1mm}\includegraphics[width=0.747\linewidth]{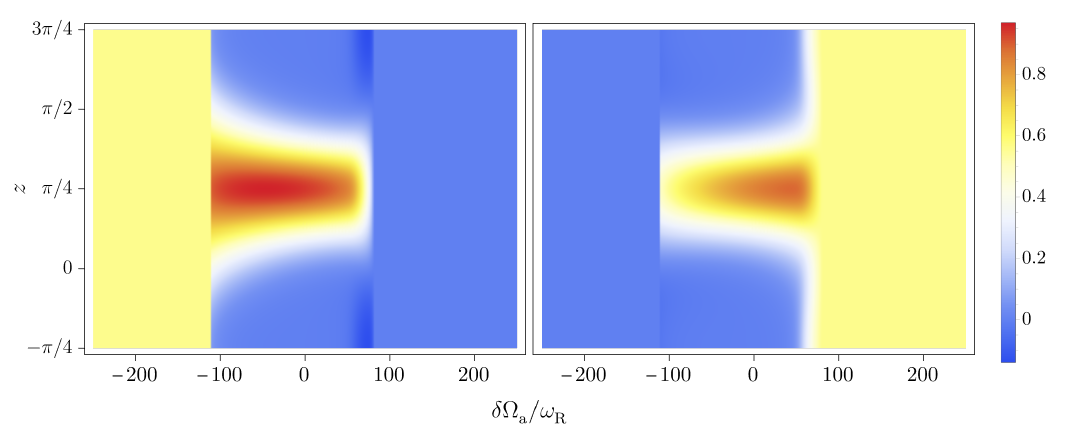}\\
\hspace*{-1mm}\includegraphics[width=0.747\linewidth]{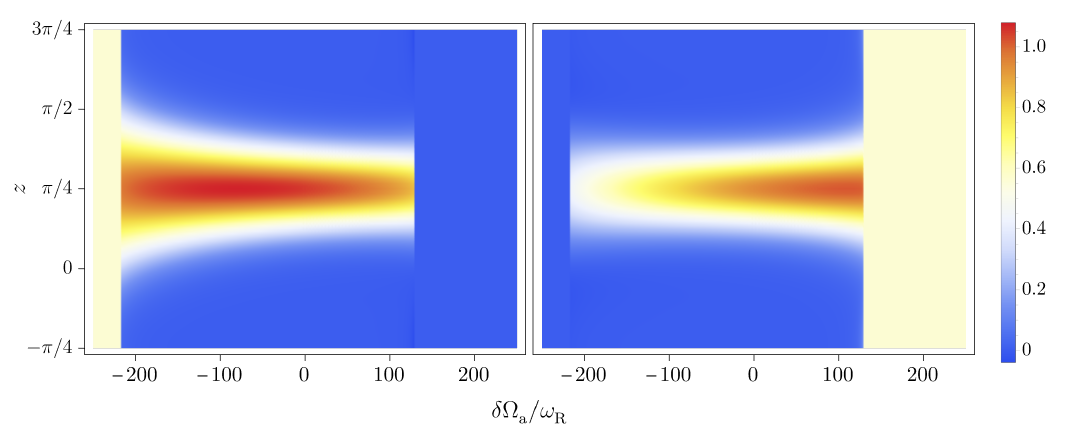}
\vspace*{-2mm}
\caption{Spatial profiles of the condensate wave functions $\psi_+^{}(z)$ (left panels) and $\psi_-^{}(z)$ (right panels) depending on the relative detuning $\delta\Omega_{\rm a}^{}$ for a fixed value of the coupling $\sqrt{N}\lambda_{\rm ex}^{}=40\omega_{\rm R}^{}$ and for three values of the coupling $\sqrt{N}\lambda\in\{20,70,120\}\;\!\omega_{\rm R}^{}$ (top, middle, bottom row). The remaining parameters are $\Omega_{\rm m}^{} = 100\omega_{\rm R}^{}$, $\Gamma_{\rm m}^{} = 10\omega_{\rm R}^{}$, $N g_{--}^{} = N g_{++}^{} = 100\omega_{\rm R}^{}$, and $N g_{+-}^{} = -90\omega_{\rm R}^{}$.}
\label{fig:4}
\end{figure*}
In Fig.~\ref{fig:3} we showcase the behavior of ${\rm Re}[\bar\alpha]$ and the population imbalance $d_{-+}^{}\equiv(N_-^{}-N_+^{})/N$ when the relative detuning is scanned across zero, for a fixed value of the coupling $\sqrt{N}\lambda_{\rm ex}^{}=40\omega_{\rm R}^{}$ and for three values of the coupling $\sqrt{N}\lambda\in\{20,70,120\}\;\!\omega_{\rm R}^{}$ (green, red, blue curves). Here, we defined $N_\pm^{}$ as the population per period of the state $|\pm\rangle$. The green curves depict a case of a continuous second-order phase transition on the negative side, followed by a continuous second-order transition on the positive side as well. This feature is manifest in both ${\rm Re}[\bar\alpha]$ and $d_{-+}^{}$. For a larger value of $\lambda$ (red curves), the transition on the negative side becomes discontinuous and of first order, while on the positive side it is still continuous. For an even larger value of $\lambda$ (blue curves), the latter transition also becomes a discontinuous first-order one.

\begin{figure*}[t]
\centering
\hspace*{\fill}
\includegraphics[width=0.36\linewidth]{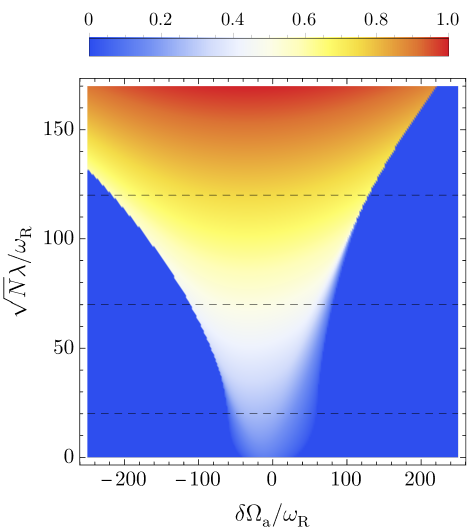}\hspace*{\fill}
\includegraphics[width=0.36\linewidth]{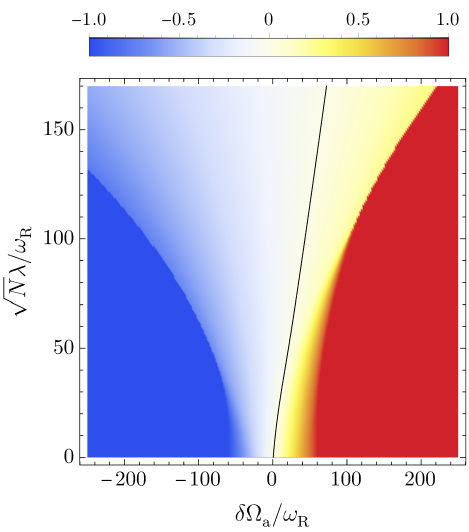}\hspace*{\fill}
\\[2mm]
\hspace*{\fill}
\includegraphics[width=0.36\linewidth]{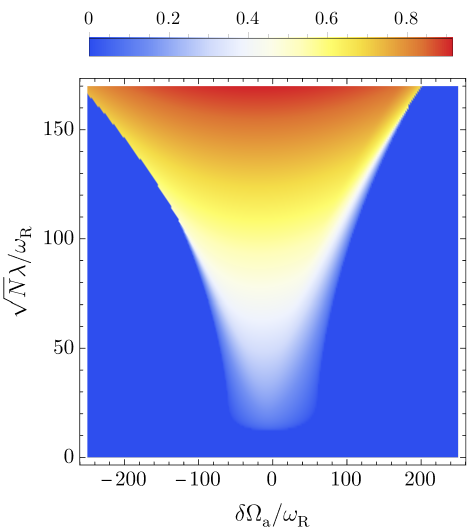}\hspace*{\fill}
\includegraphics[width=0.36\linewidth]{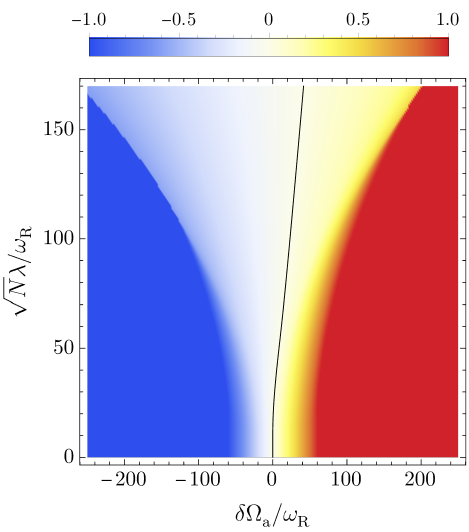}\hspace*{\fill}
\\[-2mm]
\caption{Dependence of the real part of the membrane amplitude ${\rm Re}[\bar\alpha]$ (left panels) and the population imbalance $d_{-+}^{}=(N_-^{}-N_+^{})/N$ (right panels) on the relative detuning $\delta\Omega_{\rm a}^{}$ and the coupling strength $\sqrt{N}\lambda$, for $\sqrt{N}\lambda_{\rm ex}^{}=40\omega_{\rm R}^{}$ (top row) and $\sqrt{N}\lambda_{\rm ex}^{}=25\omega_{\rm R}^{}$ (bottom row). The horizontal dashed lines in the top left panel mark those cases shown in Figs.~\ref{fig:3} and \ref{fig:4}. The black solid lines in the right panels correspond to $N_-^{}=N_+^{}$. Other parameters are: $\Omega_{\rm m}^{} = 100\omega_{\rm R}^{}$, $\Gamma_{\rm m}^{} = 10\omega_{\rm R}^{}$, $N g_{--}^{} = N g_{++}^{} = 100\omega_{\rm R}^{}$, and $N g_{+-}^{} = -90\omega_{\rm R}^{}$.}
\label{fig:5}
\end{figure*}
In order to acquire additional insight into the properties of the system, in Fig.~\ref{fig:4} we show the behavior of the condensate spatial profiles $\psi_\pm^{}(z)$ along the three curves shown in Fig.~\ref{fig:3} as we sweep the detuning from negative to positive values. One can visually distinguish the homogeneous and the self-organized phases. The wave functions $\psi_\pm^{}(z)$ are symmetric and maximal in the center of the period. The sharpness of the color gradients around the transition points reflects the (dis)continuity of the respective transition. The width of $\psi_+^{}(z)$ decreases until shortly before the second transition, while the width of $\psi_-^{}(z)$ increases monotonically. Stronger couplings $\lambda$ and $\lambda_{\rm ex}^{}$ typically lead to narrower wave function profiles due to the membrane-induced localization of the atoms.

We extend the study of the transitions discussed above to a range of values of the coupling strength $\lambda$. Figure \ref{fig:5} displays the dependence of ${\rm Re}[\bar\alpha]$ (left panels) and $d_{-+}^{}$ (right panels) on $\delta\Omega_{\rm a}^{}$ and $\lambda$ for two values of the coupling $\sqrt{N}\lambda_{\rm ex}^{}=40\omega_{\rm R}^{}$ (top row) and $\sqrt{N}\lambda_{\rm ex}^{}=25\omega_{\rm R}^{}$ (bottom row). The horizontal dashed lines in the top left panel mark explicitly those cases shown in Figs.~\ref{fig:3} and \ref{fig:4}. The single-colored blue (red) regions in the right panels correspond to the homogeneous phase where the atoms condense into the $|+\rangle$ ($|-\rangle$) state. Careful inspection of the left panels reveals that there are blue intermediate regions where the condensates are homogeneous and both atomic states have significant population. These appear for $\sqrt{N}\lambda\lesssim 20\omega_{\rm R}^{}$ and only around $|\delta\Omega_{\rm a}^{}|\approx 50\omega_{\rm R}^{}$ for the larger coupling $\lambda_{\rm ex}^{}$, as opposed to the entire range $|\delta\Omega_{\rm a}^{}|\lesssim 50\omega_{\rm R}^{}$ for the smaller coupling $\lambda_{\rm ex}^{}$. Therefore, to discriminate experimentally between the homogeneous and the inhomogeneous phase, one has first to measure the membrane displacement. If it turns out to be essentially zero, further characterization of the homogeneous phase can be done by measuring the population imbalance $d_{-+}^{}$.

The continuous nature of the transitions to the self-organized phase is manifest in the gradual changes of the color-coding on both the negative and the positive detuning side for $\sqrt{N}\lambda\lesssim 40\omega_{\rm R}^{}$ in the former case and for $\sqrt{N}\lambda\lesssim 110\omega_{\rm R}^{}$ in the latter case. In the top row, the transition on the negative side becomes sharp and discontinuous within the strip $40\omega_{\rm R}^{}\lesssim\sqrt{N}\lambda\lesssim 110\omega_{\rm R}^{}$, while the one on the positive side is still continuous. Above that, both transitions become discontinuous. For the smaller coupling $\lambda_{\rm ex}^{}$ in the bottom row, the corresponding coupling strengths are $\sqrt{N}\lambda\approx 110\omega_{\rm R}^{}$ and $\sqrt{N}\lambda\approx 160\omega_{\rm R}^{}$, respectively.
\begin{figure}[t]
\centering
\includegraphics[width=0.75\linewidth]{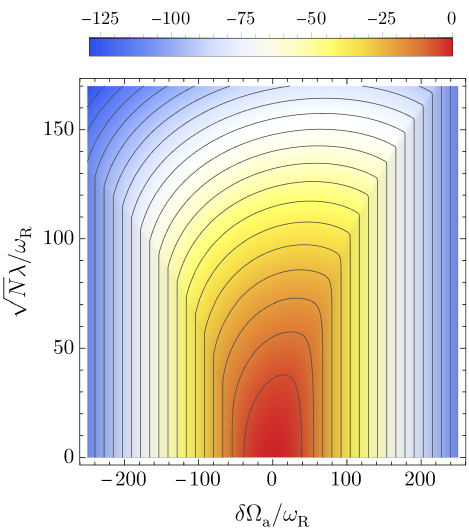}
\\[2mm]
\includegraphics[width=0.75\linewidth]{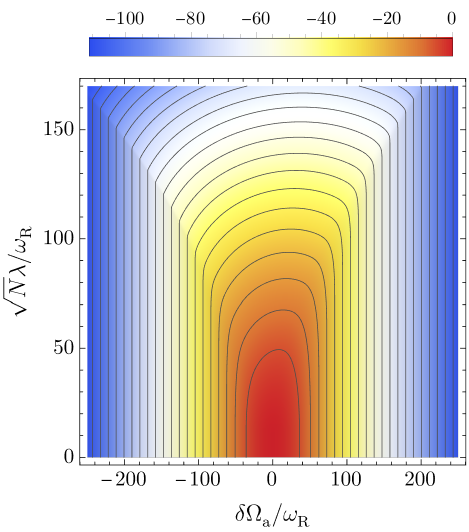}
\vspace*{-2mm}
\caption{Dependence of the energy per period, Eq.~\eqref{eq:E_pi}, on the relative detuning $\delta\Omega_{\rm a}^{}$ and the coupling strength $\sqrt{N}\lambda$, for $\sqrt{N}\lambda_{\rm ex}^{}=40\omega_{\rm R}^{}$ (top panel) and $\sqrt{N}\lambda_{\rm ex}^{}=25\omega_{\rm R}^{}$ (bottom panel). The black curves represent the contour lines. Used parameters: $\Omega_{\rm m}^{} = 100\omega_{\rm R}^{}$, $\Gamma_{\rm m}^{} = 10\omega_{\rm R}^{}$, $N g_{--}^{} = N g_{++}^{} = 100\omega_{\rm R}^{}$, and $N g_{+-}^{} = -90\omega_{\rm R}^{}$.}
\label{fig:6}
\end{figure}
\begin{figure}[t]
\centering
\includegraphics[width=0.75\linewidth]{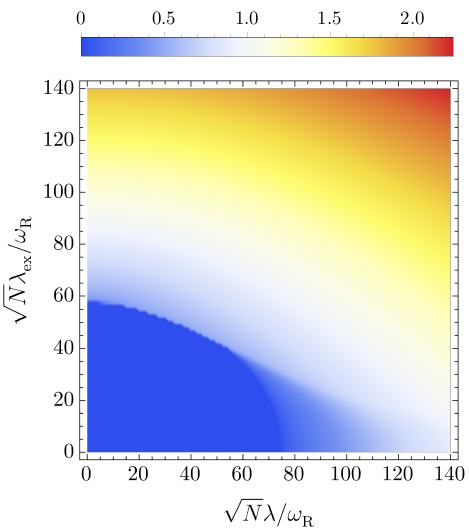}
\\[2mm]
\includegraphics[width=0.75\linewidth]{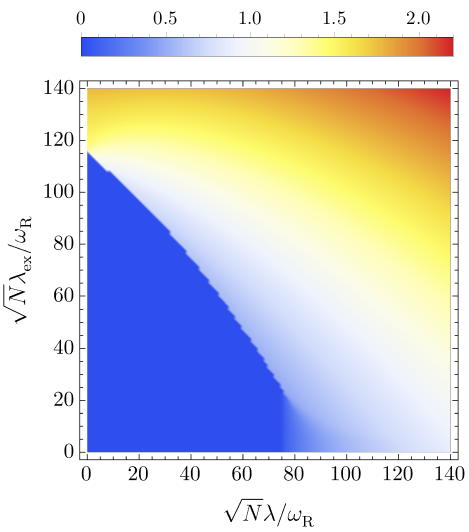}
\vspace*{-2mm}
\caption{Dependence of the real part of the membrane amplitude ${\rm Re}[\bar\alpha]$ on the coupling strengths $\sqrt{N}\lambda$ and $\sqrt{N}\lambda_{\rm ex}^{}$, for $\delta\Omega_{\rm a}^{}=-120\omega_{\rm R}^{}$ (top panel) and $\delta\Omega_{\rm a}^{}=120\omega_{\rm R}^{}$ (bottom panel). Other parameters are: $\Omega_{\rm m}^{} = 100\omega_{\rm R}^{}$, $\Gamma_{\rm m}^{} = 10\omega_{\rm R}^{}$, $N g_{--}^{} = 50\omega_{\rm R}^{}$, $N g_{++}^{} = 200\omega_{\rm R}^{}$, and $N g_{+-}^{} = -90\omega_{\rm R}^{}$.}
\label{fig:7}
\end{figure}

Further understanding can be gained by investigating the energy behavior. Figure \ref{fig:6} portrays the energy per period, given by Eq.~\eqref{eq:E_pi}, as a function of the relative detuning $\delta\Omega_{\rm a}^{}$ and the coupling strength $\sqrt{N}\lambda$, for $\sqrt{N}\lambda_{\rm ex}^{}=40\omega_{\rm R}^{}$ (top panel) and $\sqrt{N}\lambda_{\rm ex}^{}=25\omega_{\rm R}^{}$ (bottom panel). Noticeably, there are no discontinuities within the energy landscape. However, the energy features gradient jumps along the curves that correspond to discontinuous phase transitions observed in Fig.~\ref{fig:5}. This is illustrated both by the shading and the non-smoothness of the contour lines. The energy is maximal when both the relative detuning and the coupling $\lambda$ are zero. As expected, the larger either of the couplings $\lambda$ or $\lambda_{\rm ex}^{}$ is, the lower is the energy due to the stabilizing atom-membrane interaction. We also observe that for the fixed couplings the energy of the homogeneous phases decreases with increasing magnitude of $\delta\Omega_{\rm a}^{}$, while the self-ordered phase exhibits a non-monotonic dependence in this respect.

It is worth illustrating the asymmetric role played by the couplings $\lambda$ and $\lambda_{\rm ex}^{}$, which is obvious from the equations of motion \eqref{eq:MF-cond_pi}. Figure \ref{fig:7} shows ${\rm Re}[\bar\alpha]$ as a function of the two coupling strengths, for $\delta\Omega_{\rm a}^{}=-120\omega_{\rm R}^{}$ (top panel) and $\delta\Omega_{\rm a}^{}=120\omega_{\rm R}^{}$ (bottom panel). For presentation purposes, and different than in other plots, we have chosen in this context the imbalanced contact interaction strengths $N g_{--}^{} = 50\omega_{\rm R}^{}$ and $N g_{++}^{} = 200\omega_{\rm R}^{}$. The regions that correspond to the homogeneous phase are not symmetric with respect to $\lambda$ and $\lambda_{\rm ex}^{}$. For large enough relative detuning, increasing $\lambda_{\rm ex}^{}$ leads to a first-order transition for small $\lambda$ and potentially to a second-order transition for larger $\lambda$. Conversely, increasing $\lambda$ leads to a second-order transition for small $\lambda_{\rm ex}^{}$ and to a first-order transition for larger $\lambda_{\rm ex}^{}$.

\section{Discussion and outlook}\label{sec:Disscuss}

The above examples demonstrate that the considered system is versatile and offers multiple tuning knobs for changing the order of the phase transition between the homogeneous and the self-organized density wave-like state. These include, but are not limited to, the relative detuning $\delta\Omega_{\rm a}^{}$ of the atomic transition frequency $\Omega_{\rm a}^{}$ and the coupling strengths $\lambda$ and $\lambda_{\rm ex}^{}$. The two states can be experimentally distinguished by measuring the membrane displacement. To further characterize the homogeneous state, one can resort to the population imbalance $d_{-+}^{}$. We have not explored the corresponding role of the atomic contact interaction strengths $g_{\nu\nu'}^{}$ in this work. They represent, of course, further means of controlling the behavior of the system. The asymmetric nature of the atom-membrane coupling can either be further enhanced or effectively suppressed by an appropriate asymmetric choice of $g_{\nu\nu'}^{}$. Depending on the path taken through the experimental parameter space, one can encounter either first- or second-order transition hyper-surfaces. A possible future work could involve dynamics beyond the studied nonequilibrium steady state. In such a case, the system parameters could be dynamically \mbox{controlled} along different open or closed paths. In certain scenarios, hysteresis should appear and can be studied.

It would be interesting to investigate how the above mean-field analysis may be amended once quantum fluctuations have been taken into account. The spatial profiles of the two condensates critically determine which atomic normal modes are coupled to the membrane. Those that are coupled should exhibit enhanced occupation and may significantly alter primarily the homogeneous state. Therefore, the order of the observed phase transitions should be reanalyzed. Moreover, quantum fluctuations may lead to the emergence of new phases, akin to the formation of self-bound quantum droplets in Bose-Bose mixtures and dipolar Bose gases \cite{Malomed2021,Boettcher2021}. In the considered system the membrane may have an essential role to play. Beyond mean-field quantum effects may also reveal the imprint of NQPTs and their order on the statistics of the outcoupled light.

\tbl{We find it important to note the differences between the NQPTs reported here and the ones studied in Ref.~\cite{Mann2019}. First, in the cited work it was essential to have a light-induced periodic lattice in which the atoms reside. The study focused on the dynamics of the already periodically density modulated atomic cloud in this pre-existing lattice. The considered phases were characterized by oscillations of the centers of mass of the atomic cloud parts within each lattice well and around either the lattice potential minima, or the shifted equilibrium points that do not coincide with these minima. Depending on the system parameters, these NQPTs were found to be of first or second order. In the present work, the situation is quite different. Below the critical atom-membrane coupling the atomic phase is characterized by a uniform density, while above the critical coupling the self-organized density-wave-like phase spontaneously emerges, without any light-induced periodic lattice. Second, only the special case of atomic contact interaction strengths $g_{\nu\nu'}^{}=g$ was considered in Ref.~\cite{Mann2019}. Moreover, the phase transitions studied there would also occur in the absence of atom-atom interactions, i.e., for $g_{\nu\nu'}^{}=0$. Obviously, our scenario requires the overall repulsive interactions.}

In conclusion, in this paper we have presented a flexible hybrid atom-optomechanical experimental platform for the study of both continuous and discontinuous NQPTs \tbl{between the uniform and density-wave-like phases of ultracold atoms}. Importantly, it is possible to choose between the two by tuning readily accessible experimental parameters. The present work lays the foundation for further exploration of possible quantum phases of ultracold atomic gases along the lines of self-bound quantum droplets.

\begin{acknowledgments}
This work was supported by the Deutsche Forschungsgemeinschaft (DFG, German Research Foundation) via the Collaborative Research Center SFB/TR185 (Project No.~277625399) (A.P.) and via the Research Grant No.~274978739 (M.R. and M.T.). We also acknowledge the support from the DFG Cluster of Excellence ``CUI: Advanced Imaging of Matter'' $-$ EXC 2056 (Project ID No.~390715994).
\end{acknowledgments}


\begin{thebibliography}{100}

\bibitem{Sachdev2011}
S. Sachdev, {\it Quantum Phase Transitions}, 2nd ed. (Cambridge University Press, Cambridge, 2011).

\bibitem{Zinn-Justin2002}
J. Zinn-Justin, {\it Quantum Field Theory and Critical Phenomena}, 4th ed (Oxford University Press, Oxford, 2002).

\bibitem{Kleinert2001}
H. Kleinert and V. Schulte-Frohlinde, {\it Critical Properties of $\phi^4$-Theories}, (World Scientific, Singapore, 2001).

\bibitem{review1}
L. M. Sieberer, M. Buchhold, and S. Diehl,
{\it Keldysh field theory for driven open quantum systems},
\href{https://iopscience.iop.org/article/10.1088/0034-4885/79/9/096001}{Rep. Prog. Phys. {\bf 79}, 096001 (2016)}.

\bibitem{review2}
L. M. Sieberer, M. Buchhold, J. Marino, and S. Diehl,
{\it Universality in driven open quantum matter},
\href{https://doi.org/10.48550/arXiv.2312.03073}{arXiv:2312.03073 (2023)}.

\bibitem{Mitra2006}
A. Mitra, S. Takei, Y. B. Kim, and A. J. Millis,
{\it Nonequilibrium Quantum Criticality in Open Electronic Systems},
\href{https://journals.aps.org/prl/abstract/10.1103/PhysRevLett.97.236808}
{Phys. Rev. Lett. {\bf 97}, 236808 (2006)}.

\bibitem{Mitra2007}
A. Mitra and A. J. Millis,
{\it Coulomb gas on the Keldysh contour: Anderson-Yuval-Hamann representation of the nonequilibrium two-level system}, \href{https://journals.aps.org/prb/abstract/10.1103/PhysRevB.76.085342}{Phys. Rev. B {\bf 76}, 085342 (2007)}.

\bibitem{Brandes2012}
V. M. Bastidas, C. Emary, G. Schaller, and T. Brandes,
{\it Nonequilibrium quantum phase transitions in the Ising model}, \href{https://journals.aps.org/pra/abstract/10.1103/PhysRevA.86.063627}{Phys. Rev. A {\bf 86}, 063627 (2012)}.

\bibitem{Brennecke2013}
F. Brennecke, R. Mottl, K. Baumann, R. Landig, T. Donner, and T. Esslinger,
{\it Real-time observation of fluctuations at the driven-dissipative Dicke phase transition}, \href{https://www.pnas.org/doi/full/10.1073/pnas.1306993110}{Proc. Natl. Acad. Sci. USA {\bf 110}, 11763 (2013)}.

\bibitem{Marino2016}
J. Marino and S. Diehl,
{\it Driven Markovian Quantum Criticality}, \href{https://journals.aps.org/prl/abstract/10.1103/PhysRevLett.116.070407}{Phys. Rev. Lett. {\bf 116}, 070407 (2016)}.

\bibitem{Gorshkov2016}
M. F. Maghrebi and A. V. Gorshkov,
{\it Nonequilibrium many- body steady states via Keldysh formalism}, \href{https://journals.aps.org/prb/abstract/10.1103/PhysRevB.93.014307}{Phys. Rev. B {\bf 93}, 014307 (2016)}.

\bibitem{Larson2018}
J. Hannukainen and J. Larson,
{\it Dissipation-driven quantum phase transitions and symmetry breaking}, \href{https://journals.aps.org/pra/abstract/10.1103/PhysRevA.98.042113}{Phys. Rev. A {\bf 98}, 042113 (2018)}.

\bibitem{Rota2019}
R. Rota, F. Minganti, C. Ciuti, and V. Savona,
{\it Quantum Critical Regime in a Quadratically Driven Nonlinear Photonic Lattice}, \href{https://journals.aps.org/prl/abstract/10.1103/PhysRevLett.122.110405}{Phys. Rev. Lett. {\bf 122}, 110405 (2019)}.

\bibitem{Puel2019}
T. O. Puel, S. Chesi, S. Kirchner, and P. Ribeiro,
{\it Mixed-Order Symmetry-Breaking Quantum Phase Transition Far from Equilibrium}, \href{https://journals.aps.org/prl/abstract/10.1103/PhysRevLett.122.235701}{Phys. Rev. Lett. {\bf 122}, 235701 (2019)}.

\bibitem{Kennes2020}
C. Kl{\"o}ckner, C. Karrasch, and D. M. Kennes,
{\it Nonequilibrium Properties of Berezinskii-Kosterlitz-Thouless Phase Transitions}, \href{https://journals.aps.org/prl/abstract/10.1103/PhysRevLett.125.147601}{Phys. Rev. Lett. {\bf 125}, 147601 (2020)}.

\bibitem{Behrle2023}
T. Behrle, T. L. Nguyen, F. Reiter, D. Baur, B. de Neeve, M. Stadler, M. Marinelli, F. Lancellotti, S. F. Yelin, and J. P. Home, {\it Phonon Laser in the Quantum Regime}, \href{https://journals.aps.org/prl/abstract/10.1103/PhysRevLett.131.043605}{Phys. Rev. Lett. {\bf 131}, 043605 (2023)}.

\bibitem{Nagy2008}
D. Nagy, G. Szirmai and P. Domokos,
{\it Self-organization of a Bose-Einstein condensate in an optical cavity}, \href{https://link.springer.com/article/10.1140/epjd/e2008-00074-6}{Eur. Phys. J. D {\bf 48}, 127 (2008)}.

\bibitem{Ritsch2008}
C. Maschler, I. B. Mekhov, and H. Ritsch, {\it Ultracold atoms in optical lattices generated by quantized light fields}, \href{https://link.springer.com/article/10.1140/epjd/e2008-00016-4}{Eur. Phys. J. D {\bf 46}, 545 (2008)}.

\bibitem{Esslinger2010}
K. Baumann, C. Guerlin, F. Brennecke, and T. Esslinger, {\it Dicke quantum phase transition with a superfluid gas in an optical cavity}, \href{https://www.nature.com/articles/nature09009}{Nature {\bf 464}, 1301 (2010)}.

\bibitem{Ritsch2013}
H. Ritsch, P. Domokos, F. Brennecke, and T. Esslinger,
{\it Cold atoms in cavity-generated dynamical optical potentials},
\href{https://journals.aps.org/rmp/abstract/10.1103/RevModPhys.85.553}{Rev. Mod. Phys. \textbf{85}, 553 (2013)}.

\bibitem{Hemmerich2015}
J. Klinder, H. Keßler, M. Reza Bakhtiari, M. Thorwart, and A. Hemmerich, {\it Observation of a Superradiant Mott Insulator in the Dicke-Hubbard Model}, \href{https://journals.aps.org/prl/abstract/10.1103/PhysRevLett.115.230403}{Phys. Rev. Lett. {\bf 115}, 230403 (2015)}.

\bibitem{Thorwart2015}
M. Reza Bakhtiari, A. Hemmerich, H. Ritsch, and M. Thorwart, {\it Nonequilibrium Phase Transition of Interacting Bosons in an Intra-Cavity Optical Lattice}, \href{https://journals.aps.org/prl/abstract/10.1103/PhysRevLett.114.123601}{Phys. Rev. Lett. {\bf 114}, 123601 (2015)}.

\bibitem{Ritsch2021}
F. Mivehvar, F. Piazza, T. Donner, and H. Ritsch, {\it Cavity QED with quantum gases: new paradigms in many-body physics},
\href{https://doi.org/10.1080/00018732.2021.1969727}{Adv. Phys. {\bf 70}, 1 (2021)}.

\bibitem{Defenu2023}
N. Defenu, T. Donner, T. Macrì, G. Pagano, S. Ruffo, and A. Trombettoni,
{\it Long-range interacting quantum systems},
\href{https://journals.aps.org/rmp/abstract/10.1103/RevModPhys.95.035002}{Rev. Mod. Phys. \textbf{95}, 035002 (2023)}.

\bibitem{Mink}
C. D. Mink, A. Pelster, J. Benary, H. Ott, and M. Fleischhauer, {\it Variational truncated Wigner approximation for weakly interacting Bose fields: Dynamics of coupled condensates,}
\href{https://scipost.org/SciPostPhys.12.2.051}
{SciPost Phys. \textbf{12}, 051 (2022)}.

\bibitem{Ott}
J. Benary, C. Baals, E. Bernhart, J. Jiang, M. R{\"o}hrle, and H. Ott, {\it Experimental observation of a dissipative phase transition in a multi-mode many-body quantum system},
\href{https://iopscience.iop.org/article/10.1088/1367-2630/ac97b6}{New J. Phys. \textbf{24}, 103034 (2022)}.

\bibitem{Szymanska2015}
G. Dagvadorj, J. M. Fellows, S. Matyjaskiewicz, F. M. Marchetti, I. Carusotto, and M. H. Szymanska,
{\it Nonequilibrium phase transition in a two-dimensional driven open quantum system}, \href{https://journals.aps.org/prx/abstract/10.1103/PhysRevX.5.041028}{Phys. Rev. X {\bf 5}, 041028 (2015)}.

\bibitem{Szymanska2017}
A. Zamora, L. M. Sieberer, K. Dunnett, S. Diehl, and M. H. Szymanska, {\it Tuning across Universalities with a Driven Open Condensate}, \href{https://journals.aps.org/prx/abstract/10.1103/PhysRevX.7.041006}{Phys. Rev. X {\bf 7}, 041006 (2017)}.

\bibitem{Szymanska2018}
P. Comaron, G. Dagvadorj, A. Zamora, I. Carusotto, N. P. Proukakis, and M. H. Szymanska, {\it Dynamical Critical Exponents in Driven-Dissipative Quantum Systems}, \href{https://journals.aps.org/prl/abstract/10.1103/PhysRevLett.121.095302}{Phys. Rev. Lett. {\bf 121}, 095302 (2018)}.

\bibitem{Nyman2018}
H. J. Hesten, R. A. Nyman, and F. Mintert,
{\it Decondensation in Nonequilibrium Photonic Condensates: When Less Is More},
\href{https://journals.aps.org/prl/abstract/10.1103/PhysRevLett.120.040601}{Phys. Rev. Lett. \textbf{120}, 040601 (2018)}.

\bibitem{Nyman2021}
J. D. Rodrigues, H. S. Dhar, B. T. Walker, J. M. Smith, R. F. Oulton, F. Mintert, and R. A. Nyman, {\it Learning the Fuzzy Phases of Small Photonic Condensates},
\href{https://journals.aps.org/prl/abstract/10.1103/PhysRevLett.126.150602}{Phys. Rev. Lett. \textbf{126}, 150602 (2021)}.

\bibitem{Weitz}
F. {\"O}zt{\"u}rk, T. Lappe, G. Hellmann, J. Schmitt, J. Klaers, F. Vewinger, J. Kroha, and M. Weitz, {\it Observation of a non-Hermitian phase transition in an optical quantum gas},
\href{https://www.science.org/doi/10.1126/science.abe9869?ijkey=E3tj1dY9WeEEg&keytype=ref&siteid=sci+}{Science \textbf{372}, 88 (2021)}.

\bibitem{Heyl}
M. Heyl, {\it Dynamical quantum phase transitions},
\href{https://iopscience.iop.org/article/10.1088/1361-6633/aaaf9a}{Rep. Prog. Phys. \textbf{81}, 054001 (2018)}.

\bibitem{Widera}
L.-N. Wu, J. Nettersheim, J. Fe{\ss}, A. Schnell, S. Burgardt, S. Hiebel, D. Adam, A. Eckardt, and A. Widera, {\it Indication of critical scaling in time during the relaxation of an open quantum system},
\href{https://www.nature.com/articles/s41467-024-46054-9}{Nature Comm. \textbf{15}, 1714 (2024)}.

\bibitem{Treutlein2013}
B. Vogell, K. Stannigel, P. Zoller, K. Hammerer, M. T. Rakher, M. Korppi, A. J{\"o}ckel, and P. Treutlein, {\it Cavity-enhanced long-distance coupling of an atomic ensemble to a micromechanical membrane}, \href{https://journals.aps.org/pra/abstract/10.1103/PhysRevA.87.023816}{Phys. Rev. A {\bf 87}, 023816 (2013)}.

\bibitem{Vogell2015}
B. Vogell, T. Kampschulte, M. T. Rakher, A. Faber, P. Treutlein, K. Hammerer, and P. Zoller, {\it Long distance coupling of a quantum mechanical oscillator to the internal states of an atomic ensemble}, \href{https://iopscience.iop.org/article/10.1088/1367-2630/17/4/043044/meta}{New J. Phys. {\bf 17}, 043044 (2015)}.

\bibitem{Treutlein2015}
A. J{\"o}ckel, A. Faber, T. Kampschulte, M. Korppi, M. T. Rakher, and P. Treutlein, {\it Sympathetic cooling of a membrane oscillator in a hybrid mechanical–atomic system}, \href{https://www.nature.com/articles/nnano.2014.278}{Nat. Nanotechnol. {\bf 10}, 55 (2015)}.

\bibitem{Zhong2017}
H. Zhong, G. Fl{\"a}schner, A. Schwarz, R. Wiesendanger, P. Christoph, T. Wagner, A. Bick, C. Staarmann, B. Abeln, K. Sengstock, and C. Becker {\it A millikelvin all-fiber cavity optomechanical apparatus for merging with ultra-cold atoms in a hybrid quantum system}, \href{https://pubs.aip.org/aip/rsi/article-abstract/88/2/023115/358324/A-millikelvin-all-fiber-cavity-optomechanical?redirectedFrom=fulltext}{Rev. Sci. Instrum. {\bf 88}, 023115 (2017)}.

\bibitem{Becker2018}
P. Christoph, T. Wagner, H. Zhong, R. Wiesendanger, K. Sengstock, A. Schwarz, and C. Becker, {\it Combined feedback and sympathetic cooling of a mechanical oscillator coupled to ultracold atoms}, \href{https://iopscience.iop.org/article/10.1088/1367-2630/aadf20}{New J. Phys. {\bf 20}, 093020 (2018)}.

\bibitem{Treutlein2018}
A. Vochezer, T. Kampschulte, K. Hammerer, and P. Treutlein, {\it Light-Mediated Collective Atomic Motion in an Optical Lattice Coupled to a Membrane}, \href{https://journals.aps.org/prl/abstract/10.1103/PhysRevLett.120.073602}{Phys. Rev. Lett. {\bf 120}, 073602 (2018)}.

\bibitem{Mann2018}
N. Mann, M. Reza Bakhtiari, A. Pelster, and M. Thorwart, {\it Nonequilibrium Quantum Phase Transition in a Hybrid Atom-Optomechanical System}, \href{https://journals.aps.org/prl/abstract/10.1103/PhysRevLett.120.063605}{Phys. Rev. Lett. {\bf 120}, 063605 (2018)}.

\bibitem{Gao2019}
C. Gao and Z. Liang, {\it Steady-state phase diagram of quantum gases in a lattice coupled to a membrane}, \href{https://journals.aps.org/pra/abstract/10.1103/PhysRevA.99.013629}{Phys. Rev. A {\bf 99}, 013629 (2019)}.

\bibitem{Treutlein2022b}
T. M. Karg, B. Gouraud, C. T. Ngai, G.-L. Schmid, K. Hammerer, and P. Treutlein,
{\it Light-mediated strong coupling between a mechanical oscillator and atomic spins 1 meter apart},
\href{https://www.science.org/doi/10.1126/science.abb0328}{Science \textbf{369}, 174 (2020)}.

\bibitem{Treutlein2022}
G.-L. Schmid, C. T. Ngai, M. Ernzer, M. B. Aguilera, T. M. Karg, and P. Treutlein,
{\it Coherent Feedback Cooling of a Nanomechanical Membrane with Atomic Spins}, \href{https://journals.aps.org/prx/abstract/10.1103/PhysRevX.12.011020}{Phys. Rev. X {\bf 12}, 011020 (2022)}.

\bibitem{Treutlein2023}
M. Ernzer, M. B. Aguilera, M. Brunelli, G.-L. Schmid, T. M. Karg, C. Bruder, P. P. Potts, and P. Treutlein,
{\it Optical Coherent Feedback Control of a Mechanical Oscillator}, \href{https://journals.aps.org/prx/abstract/10.1103/PhysRevX.13.021023}{Phys. Rev. X {\bf 13}, 021023 (2023)}.

\bibitem{Hammerer2009}
K. Hammerer, M. Aspelmeyer, E. S. Polzik, and P. Zoller, {\it Establishing Einstein-Poldosky-Rosen Channels between Nanomechanics and Atomic Ensembles}, \href{https://journals.aps.org/prl/abstract/10.1103/PhysRevLett.102.020501}{Phys. Rev. Lett. {\bf 102}, 020501 (2009)}.

\bibitem{Kimble2009}
K. Hammerer, M. Wallquist, C. Genes, M. Ludwig, F. Marquardt, P. Treutlein, P. Zoller, J. Ye, and H. J. Kimble, {\it Strong Coupling of a Mechanical Oscillator and a Single Atom}, \href{https://journals.aps.org/prl/abstract/10.1103/PhysRevLett.103.063005}{Phys. Rev. Lett. {\bf 103}, 063005 (2009)}.

\bibitem{Wallquist2009}
M. Wallquist, K. Hammerer, P. Rabl, M. Lukin, and P. Zoller, {\it Hybrid quantum devices and quantum engineering}, \href{https://iopscience.iop.org/article/10.1088/0031-8949/2009/T137/014001}{Phys. Scr. {\bf T137}, 014001 (2009)}.

\bibitem{Palma2010}
M. Paternostro, G. De Chiara, and G. M. Palma, {\it Cold-Atom-Induced Control of an Optomechanical Device}, \href{https://journals.aps.org/prl/abstract/10.1103/PhysRevLett.104.243602}{Phys. Rev. Lett. {\bf 104}, 243602 (2010)}.

\bibitem{Genes2011}
C. Genes, H. Ritsch, M. Drewsen, and A. Dantan, {\it Atom-membrane cooling and entanglement using cavity electromagnetically induced transparency}, \href{https://journals.aps.org/pra/abstract/10.1103/PhysRevA.84.051801}{Phys. Rev. A {\bf 84}, 051801(R) (2011)}.

\bibitem{Lau2018}
H. K. Lau, A. Eisfeld, and J. M. Rost, {\it Cavity-free quantum optomechanical cooling by atom-modulated radiation}, \href{https://journals.aps.org/pra/abstract/10.1103/PhysRevA.98.043827}{Phys. Rev. A {\bf 98}, 043827 (2018)}.

\bibitem{Ockeloen2013}
C. F. Ockeloen, R. Schmied, M. F. Riedel, and P. Treutlein, {\it Quantum Metrology with a Scanning Probe Atom Interferometer}, \href{https://journals.aps.org/prl/abstract/10.1103/PhysRevLett.111.143001}{Phys. Rev. Let. {\bf 111}, 143001 (2013)}.

\bibitem{Polzik2015}
E. S. Polzik and K. Hammerer, {\it Trajectories without quantum uncertainties}, \href{https://onlinelibrary.wiley.com/doi/10.1002/andp.201400099}{Ann. Phys. (Leipzig) {\bf 527}, A15 (2015)}.

\bibitem{Polzik2017}
C. B. Møller, R. A. Thomas, G. Vasilakis, E. Zeuthen, Y. Tsaturyan, M. Balabas, K. Jensen, A. Schliesser, K. Hammerer, and E. S. Polzik, {\it Quantum back-action-evading measurement of motion in a negative mass reference frame}, \href{https://www.nature.com/articles/nature22980}{Nature {\bf 547}, 191 (2017)}.

\bibitem{Mann2019}
N. Mann, A. Pelster, and M. Thorwart, {\it Tuning the order of the nonequilibrium quantum phase transition in a hybrid atom-optomechanical system}, \href{https://iopscience.iop.org/article/10.1088/1367-2630/ab51fa}{New J. Phys. {\bf 21}, 113037 (2019)}.

\bibitem{Zhou2021}
J.-M. Cheng, Z.-W. Zhou, G.-C. Guo, H. Pu, and X.-F. Zhou, {\it Bose-Einstein condensates in an atom-optomechanical system with effective global nonuniform interaction}, \href{https://journals.aps.org/pra/abstract/10.1103/PhysRevA.103.023328}{Phys. Rev. A {\bf 103}, 023328 (2021)}.

\bibitem{Zoller2004}
C. W. Gardiner and P. Zoller, {\it Quantum Noise: A Handbook of Markovian and Non-Markovian Quantum Stochastic Methods with Applications to Quantum Optics}, Springer Series in Synergetics (Springer, Berlin, 2004).

\bibitem{Malomed2021}
Z.-H. Luo, W. Pang, B. Liu, Y.-Y. Li, B. A. Malomed,
{\it A new form of liquid matter: Quantum droplets}, \href{https://link.springer.com/article/10.1007/s11467-020-1020-2}{Front. Phys. {\bf 16}, 32201 (2021)}.

\bibitem{Boettcher2021}
F. B{\"o}ttcher, J.-N. Schmidt, J. Hertkorn, K. S. H. Ng, S. D. Graham, M. Guo, T. Langen, and T. Pfau,
{\it New states of matter with fine-tuned interactions: quantum droplets and dipolar supersolids}, \href{https://iopscience.iop.org/article/10.1088/1361-6633/abc9ab}{Rep. Prog. Phys. {\bf 84}, 012403 (2021)}.

\end{thebibliography}
\end{document}